\documentclass[sigconf, authorversion,nonacm]{acmart}
\AtBeginDocument{%
  \providecommand\BibTeX{{%
    \normalfont B\kern-0.5em{\scshape i\kern-0.25em b}\kern-0.8em\TeX}}}

\usepackage{enumitem, xcolor, soul}
\usepackage{multirow}
\begin{document}

\title{Transductive Spiking Graph Neural Networks for Loihi}

\author{Shay Snyder}
\authornote{These authors contributed equally to this work}
\email{ssnyde9@gmu.edu}
\orcid{0000-0002-3369-3478}
\affiliation{%
  \institution{George Mason University}
  \streetaddress{4400 University Dr}
  \city{Fairfax}
  \state{Virginia}
  \country{USA}
  \postcode{22030}
}
\author{Victoria Clerico}
\authornotemark[1]
\email{mclerico@gmu.edu}
\orcid{0009-0005-2383-0335}
\affiliation{%
  \institution{George Mason University}
  \streetaddress{4400 University Dr}
  \city{Fairfax}
  \state{Virginia}
  \country{USA}
  \postcode{22030}
}
\author{Guojing Cong}
\email{congg@ornl.gov}
\orcid{0000-0003-0850-7714}
\affiliation{%
  \institution{Oak Ridge National Laboratory}
  \city{Oak Ridge}
  \state{Tennessee}
  \country{USA}
  \postcode{37830}
}
\author{Shruti Kulkarni}
\email{kulkarnisr@ornl.gov}
\orcid{0000-0001-6894-9851}
\affiliation{%
  \institution{Oak Ridge National Laboratory}
  \city{Oak Ridge}
  \state{Tennessee}
  \country{USA}
  \postcode{37830}
}
\author{Catherine Schuman}
\email{cschuman@utk.edu}
\orcid{0000-0001-6894-9851}
\affiliation{%
  \institution{University of Tennessee - Knoxville}
  \city{Knoxville}
  \state{Tennessee}
  \country{USA}
  \postcode{37996}
}
\author{Sumedh R. Risbud}
\email{sumedh.risbud@intel.com}
\orcid{0000-0003-4777-1139}
\affiliation{%
  \institution{Intel Labs}
  \streetaddress{2200 Mission College Blvd}
  \city{Santa Clara}
  \state{California}
  \country{USA}
  \postcode{95054}
}
\author{Maryam Parsa}
\email{mparsa@gmu.edu}
\orcid{0000-0002-4855-4593}
\affiliation{%
  \institution{George Mason University}
  \streetaddress{4400 University Dr}
  \city{Fairfax}
  \state{Virginia}
  \country{USA}
  \postcode{22030}
}

\renewcommand{\shortauthors}{Snyder, Clerico, et al.}

\begin{abstract}
    Graph neural networks have emerged as a specialized branch of deep learning, designed to address problems where pairwise relations between objects are crucial. Recent advancements utilize graph convolutional neural networks to extract features within graph structures. Despite promising results, these methods face challenges in real-world applications due to sparse features, resulting in inefficient resource utilization. Recent studies draw inspiration from the mammalian brain and employ spiking neural networks to model and learn graph structures. However, these approaches are limited to traditional Von Neumann-based computing systems, which still face hardware inefficiencies. In this study, we present a fully neuromorphic implementation of spiking graph neural networks designed for Loihi 2. We optimize network parameters using Lava Bayesian Optimization, a novel hyperparameter optimization system compatible with neuromorphic computing architectures. We showcase the performance benefits of combining neuromorphic Bayesian optimization with our approach for citation graph classification using fixed-precision spiking neurons. Our results demonstrate the capability of integer-precision, Loihi 2 compatible spiking neural networks in performing citation graph classification with comparable accuracy to existing floating point implementations.
\end{abstract}






\maketitle

\section{Introduction}
Inspired by its success in image processing, researchers have recently adapted the concept of convolution to analyze the relationships within graph structures. This technique is termed graph convolution~\cite{zhang2019graph}. Compared to traditional convolutional neural networks that extract features from local receptive fields~\cite{lecun1995convolutional}, graph convolutional neural networks (GCNN) extract information from heterogeneous graph structures by understanding the relationship between interconnected nodes~\cite{zhang2019graph}. This learning approach has been successfully applied to a variety of real-world machine learning applications such as material design~\cite{Xie_2018}, text classification~\cite{yao2018graph}, citation graph classification~\cite{cong2022semi}, and pharmaceutical discovery~\cite{10.1093/bib/bbz042}.

Graph neural networks are particularly effective in applications with high-levels of sparsity by employing semi-supervised learning~\cite{semisupervised, semisupervised2, semisupervised4}. Popular benchmarks such as Sen et al.~\cite{Sen_Namata_Bilgic_Getoor_Galligher_Eliassi-Rad_2008} achieved high accuracy despite having only 6\% of the graph nodes labeled as training. While existing approaches can achieve high accuracy, they often struggle to efficiently model real-world graphs. These data structures are characterized by a mixture of highly interconnected regions and immensely sparse areas. Popular hardware accelerators, such as graphic processing units, struggle to efficiently utilize computational resources when processing irregularly clustered computational graphs since their design is optimized for dense workloads.

Neuromorphic computing and biologically inspired networks have emerged as prevalent research fields to address sparse data representations. Recent works demonstrate the performance of spiking neural networks for a variety of graph learning tasks such as the minimum spanning tree~\cite{graph_neuro3} and shortest path problems~\cite{graph_neuro1, graph_neuro2}. These learning approaches posses unique properties making them especially attractive for implementing graph learning algorithms through event-driven communication between individual computational units (spiking neurons). Moreover, research investment from companies such as Intel~\cite{loihi} and IBM~\cite{7229264} has lead to the development of ASIC processors specifically designed to perform these neuromorphic computations.

Existing applications of GCNNs extract hidden features within the network, where the convolution operation is performed over adjacent nodes and information from neighboring nodes is iteratively propagated and aggregated~\cite{benchmark4}. Similarly, other methods employ graph attention over the neighborhood of each node and learn graph embeddings~\cite{benchmark2}. While existing methods utilize a separate neural network for processing the graph, our approach builds the network structure directly based on graph anatomy. This transforms learning into a transductive approach where learning happens within the citation graph itself. Additionally, existing methods using spiking neural networks for transductive learning have been limited to simulated, floating point environments that are not natively compatible with neuromorphic hardware~\cite{cong2023, icons22guojing}. This paper investigates the capability of fixed-precision spiking neural networks designed for Intel's neuromorphic processor, Loihi 2~\cite{loihi}.

We introduce a new, fixed-precision neuron model (extending from the traditional leaky-integrate-and-fire (LIF) known as \textbf{LIF Long Reset}. This neuron model allows the network to fully reset at predefined intervals by setting all currents and voltages to zero for multiple timesteps. 
In tandem with Lava Bayesian Optimization~\cite{10.1145/3589737.3605998}, we optimize network parameters to achieve performance approaching existing floating point methods while limiting intra-network calculations to integer precision. The key contributions of this work are summarized as follows:

\begin{itemize}[leftmargin=6mm]
    \item We provide a novel fixed precision spiking neural network model for transductive graph learning in Lava.
    \item We introduce the LIF Long Reset neuron model that allows the network to self-stabilize at pre-defined intervals while also contributing to the Lava software framework.
    \item Our integer-precision implementation in Lava exhibits similar performance to an existing floating point implementation in NEST~\cite{cong2023}.
\end{itemize}

 
     

\section{Research Methods}
Citation network graphs consist of interconnected paper nodes based on their citations. Each node represents a distinct research paper, while edges signify citations between papers. Moreover, each paper is labeled with a specific paper topic. The citation graph from Cora~\cite{cora} includes 2708 paper nodes, 5258 edges, and 7 topics: rule learning, neural networks, case-based, genetic algorithms, theory, reinforcement learning, and probabilistic methods. Many citation network datasets also include a series of features linked to each paper. Our current implementation does not include this additional information.

We randomly select 20 nodes per paper topic to serve as the labeled training neurons. The remaining paper nodes are proportionally split into validation and testing neurons. This is represented in our spiking neural network by creating three individual clusters of spiking neurons; 140 for training, 140 for validation, and 2428 for testing. Lastly, we initialize another cluster of neurons representing the topic nodes with 7 neurons.

The paper neurons are initialized with a voltage decay of 0 and a voltage threshold equal to 1. For each citation (edge) in the citation graph from paper \textbf{A} to paper \textbf{B}, we create a corresponding synapse with an initial weight of 100. These synapses from paper \textbf{A} to paper \textbf{B} can occur within the same neuron cluster or interconnect neurons from different clusters. The process is repeated in the reverse direction from paper \textbf{B} to paper \textbf{A}, encouraging recurrent communication. Each paper neuron within the training cluster is directly connected to the correct topic neuron with a weight of 1 along with an identical reverse connection.

\begin{figure}[h]
\centering
\includegraphics[width=0.3\textwidth]{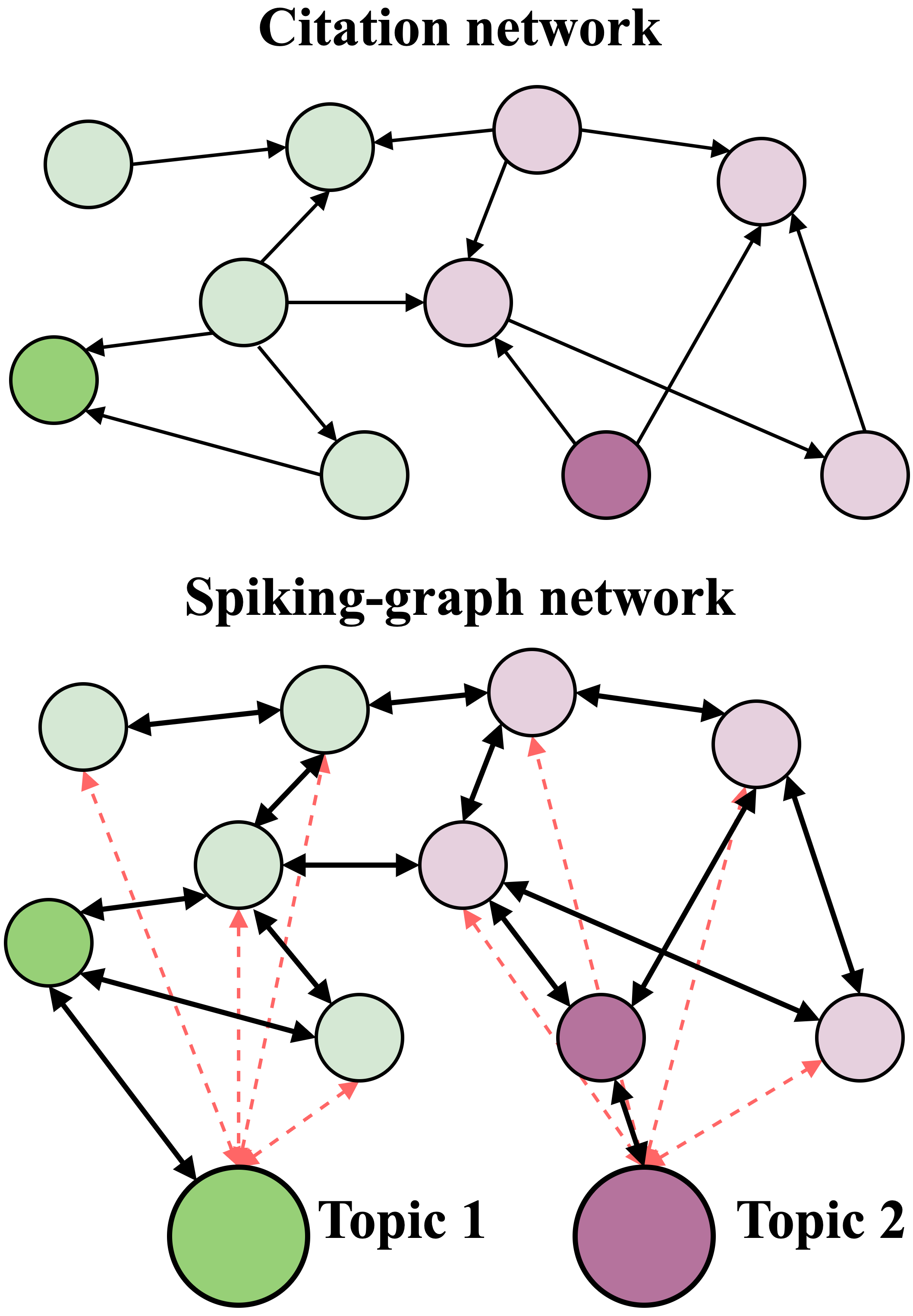}
\caption{A simplified graph demonstrating the architecture of our spiking neural network for citation graph classification. Red dotted arrows represent the learnable synapses.}
\label{fig:citation-graph}
\end{figure}

The testing and validation neurons have dense connectivity with all topic neurons and learnable synaptic weights. This is demonstrated in Figure~\ref{fig:citation-graph}. Using Hebbian learning rules~\cite{Gerstner1996stdp}, we employ bi-directional spike timing-dependent plasticity (STDP) to learn the correct paper-to-topic connection for each paper in the testing and validation sets.

STDP-based learning uses the temporal correlation between spikes registered in time $t_{pre}$ from pre-synaptic neurons and spikes registered at $t_{post}$ from post-synaptic neurons to perform localized learning.
This mechanism strengthens the synaptic weight between neurons when the pre-synaptic neuron fires before the post-synaptic neuron and weakens the connection if the pre-synaptic neuron fires afterwords. These connections all begin with an initial weight of 0. The learning rule is configured with a learning rate of $2$, $a_-=-1$, $a_+=1$, $\tau_-=30$, and $\tau_+=30$\footnote{See Gerstner et al. \cite{Gerstner1996stdp} for more information on STDP parameters}. These parameters are also summarized in Table~\ref{tab:params}.

To classify any given paper, we send a single spike at $t=0$ to the corresponding paper neuron and allow it to propagate through the network for $t_s$ time steps.
At the end of this interval, we extract the predicted label by interpreting information from the trained graph. One strategy is to infer the learned topic as the topic neuron with the highest firing rate. Another approach is to choose the predicted label based on the learned synaptic weights between the specific paper and all topic neurons. To avoid encouraging increased synaptic activity, we employ the second method where the label is inferred via the corresponding topic to paper synaptic weights.
With the topic (pre-synaptic) neurons firing after the paper (post-synaptic) neurons, the lowest weight (largest negative magnitude) is selected because STDP depresses the synaptic weight between neurons when the post-synaptic neuron fires before the pre-synaptic neuron.

\begin{figure}[h]
\centering
\includegraphics[width=0.4\textwidth]{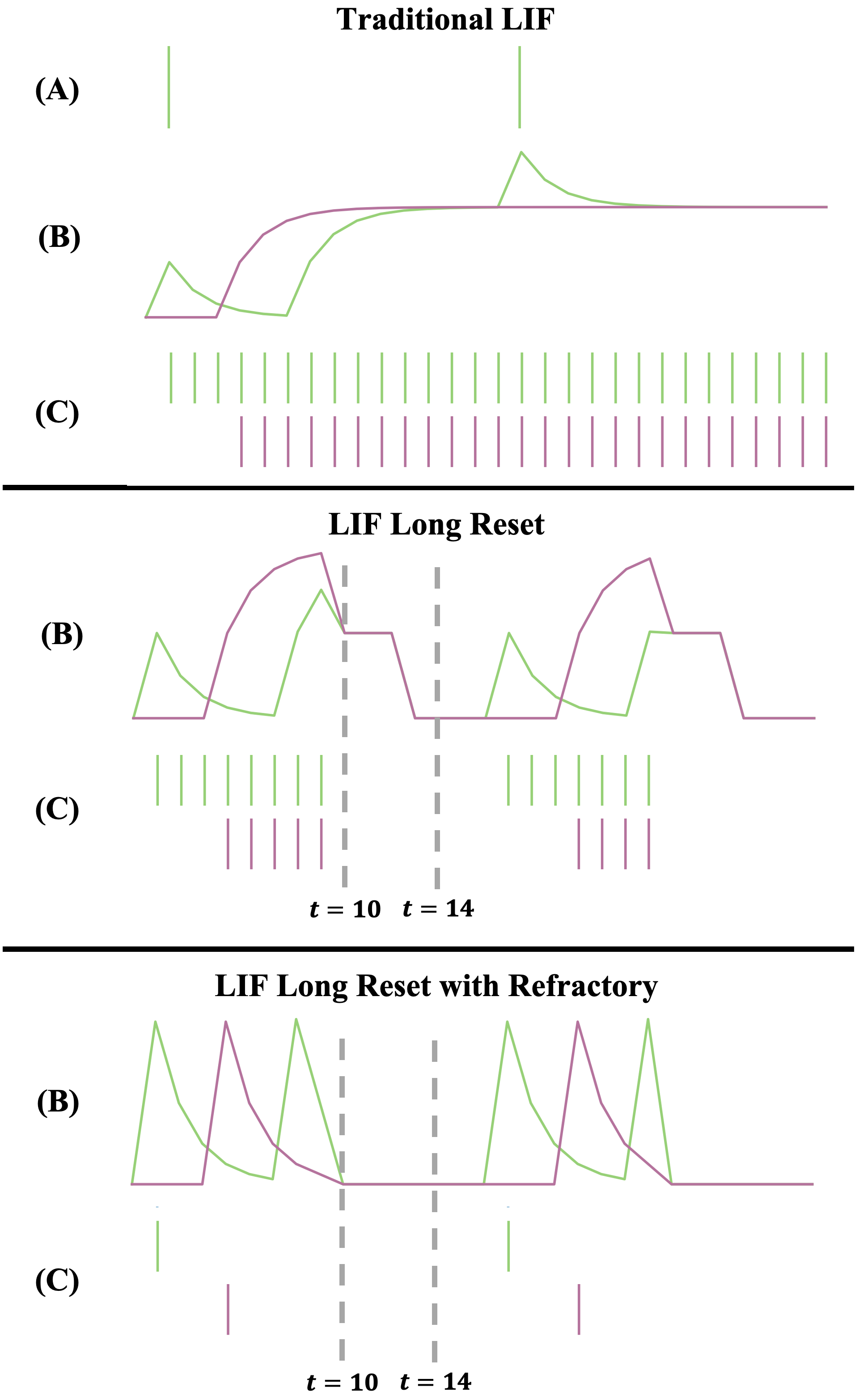}
\caption{The inter-neuron dynamics of our LIF Long Reset neuron model compared to a traditional LIF neuron through two neurons recurrently connected with fully-connected layers. (A) Input spikes from the ring buffer (B) Synaptic current (C) Post synaptic spike rasters from both neurons. Each neuron was initialized with a current decay of 0.5. Each LIF Long Reset neuron variant was configured with a reset interval of 10 time steps with a reset period of 4 time steps.}
\label{fig:neuron-dynamics}
\end{figure}

To reset our dynamic spiking graph between paper evaluations, we created a new neuron based on the traditional leaky-integrate-and-fire (LIF) model, called the \textbf{LIF Long Reset} neuron. This neuron can intermittently stop firing at pre-specified reset intervals $\mathbf{T_r}$ for a given reset length. We constrain the reset interval $T_r$ to be equal to the number of spike propagation steps such that $T_r = T_s$. This allows us to accurately decode the learned weights before the network is allowed to reach homeostasis. Our model also supports refractory dynamics so it will only fire once for any given paper evaluation. The performance implications and inter-neuron characteristics for this model are highlighted in Figure~\ref{fig:neuron-dynamics}.

\begin{figure*}[t]
\centering
\includegraphics[width=0.85\textwidth]{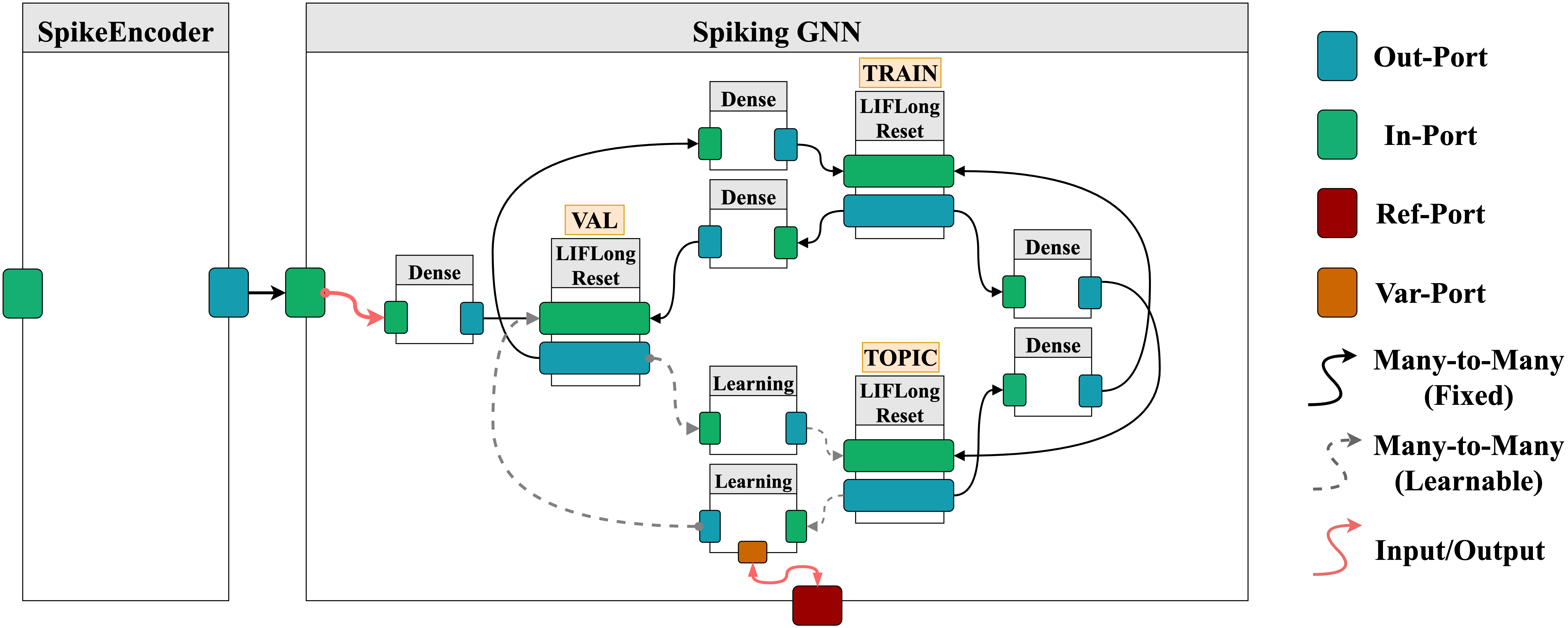}
\caption{The multi-process and subprocess model architecture representing our Spiking GNN approach for citation network classification in Lava. For simplicity, connections within the validation and testing neurons clusters along with the entire testing cluster have been omitted. \textit{Ref and Var Ports} allow for read and write operations between interconnected processes.}
\label{fig:process-architecture}
\end{figure*}

Developing and deploying spiking neural networks on Loihi 2 requires prototyping the model using Intel's neuromorphic framework Lava~\cite{lava}. The fundamental blocks of Lava are \textit{Processes}, which operate in parallel and asynchronously with each other. Lava's \textit{Processes} have their own private memory and communicate with other processes through channel-based message passing. \textit{Processes} can be used to implement a single neuron, a cluster of neurons, or an entire neural network at the complexity level of ResNet~\cite{lava}. A \textit{Process} is defined by \textit{input ports}, \textit{internal variables} and \textit{output ports}. The \textit{Processes} only define an interface with state variables and communication ports, therefore they do not provide a behavioral implementation. As such, the behavior is implemented in \textit{Process Models}\footnote{See \url{http://lava-nc.org} for details about Lava concepts like \textit{Process} and \textit{Process Models}}. 

Our learning pipeline consists of one process for generating input spikes (\textit{Spike Encoder}) and a subprocess model for the spiking graph neural network (\textit{Spiking GNN}), see Figure~\ref{fig:process-architecture}. The \textit{Spike Encoder} process is responsible for sending the single spike through the paper neuron under evaluation.
The \textit{Spiking GNN} subprocess model synchronously runs three types of processes: (1) \textit{LIFLongReset} for the paper and topic neurons, (2) \textit{Dense} for the fixed synaptic connections and (3) \textit{Learning Dense} for the learnable synapses of the graph, see Figure \ref{fig:process-architecture}. The learning rule for the learnable synapses is STDP-based (\textit{STDPLoihi}) and is implemented as described in~\cite{Gerstner1996stdp}.

At the end of execution, classification accuracy is computed outside the process by taking the predicted labels for each paper and comparing with the ground truth. With this multi-process architecture in Lava, our spiking graph has full support for simulation with \textit{fixed-precision, integer computations} as required by Loihi 2.

Each of our spiking neuron clusters and instances of STDP have numerous parameters such as current decay, voltage decay, learning rate, and synaptic delays. With our architecture consisting of multiple clusters of neurons and multiple instances of STDP, it is extremely difficult to manually specify each individual parameter.
To automate this process, we integrate our event-driven process architecture with Lava Bayesian Optimization (Lava BO)~\cite{10.1145/3589737.3605998}. This framework provides a native, event-based framework for optimizing neuromorphic systems or other event-driven technologies developed in Lava.

\begin{figure}
    \centering
    \includegraphics[width=0.45\textwidth]{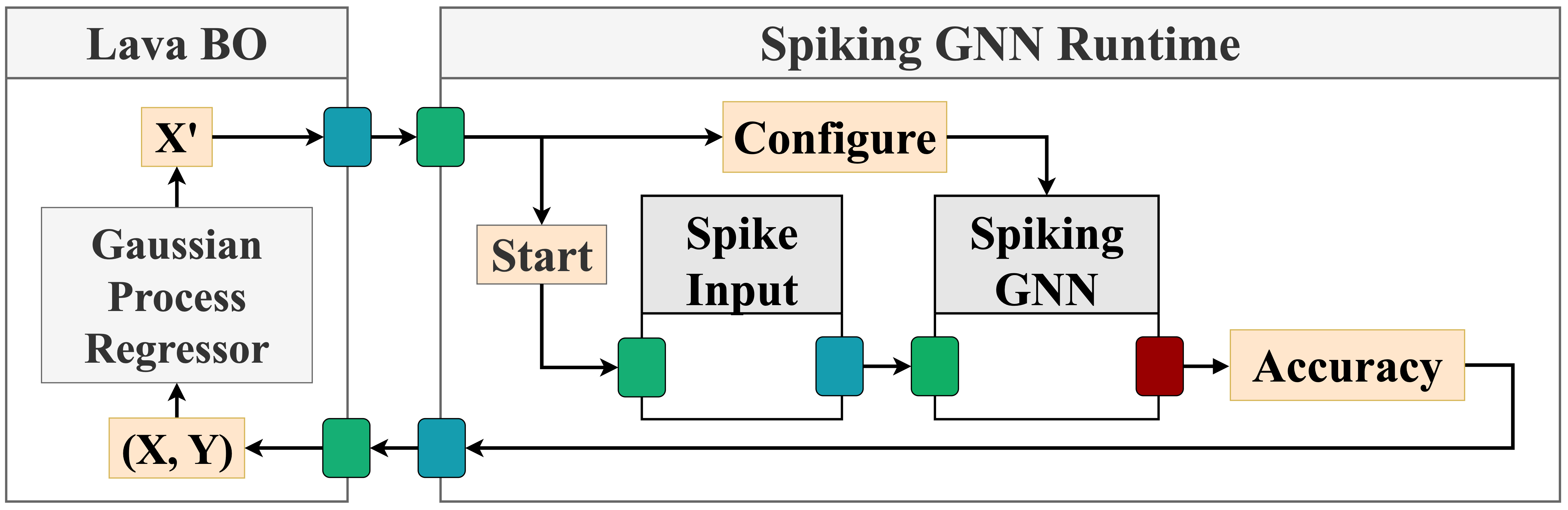}
    \caption{The runtime control flow provided by the Spiking GNN Runtime process whenever it receives new parameters from LavaBO~\cite{10.1145/3589737.3605998}. For more details on the specifics of the Spiking GNN, see Figure~\ref{fig:process-architecture}.}
    \label{fig:bo-gnn}
\end{figure}


We develop a wrapper process, known as the \textit{Spiking GNN Runtime}, to integrate Lava BO with our spiking graph neural network. As shown in Figure~\ref{fig:bo-gnn}, this process receives a vector of hyperparameters from Lava BO, configures the GNN accordingly, and evaluates its performance across the entire validation set. Once the accuracy is determined, both the parameters and the resulting accuracy value are returned to Lava BO~\cite{10.1145/3589737.3605998}. This iterative process continues until a predefined maximum number of iterations is reached.
\section{Results \& Discussion}
With our fixed-precision spiking graph neural in Lava, we perform citation network classification with the Cora dataset~\cite{cora}. We compare our approach against a floating-point implementation developed in NEST~\cite{cong2023, Gewaltig:NEST}. This comparison highlights the performance differences between fixed-precision, Loihi 2~\cite{loihi} compatible architectures compared against traditional floating point implementations that lack native support for neuromorphic hardware.
Without this support, networks developed in NEST cannot be easily validated and designed for edge applications where neuromorphic computing systems show great promise over traditional artificial neural networks. 

On the other hand, Lava is designed as a prototyping tool for deployment to heterogeneous, event-based architectures such as Loihi 2. The Lava runtime environment executes multiple asynchronous processes with channel-based communication, providing a fine-grained approach to building asynchronous systems. This event-based communication is ideal for scenarios where independent processes need to exchange data without blocking. This framework is developed in Python, which hurdles in fully exploiting the potential for concurrent processing. Python's interpreter restricts multiple processes from utilizing different CPU cores simultaneously. However, this parallelism and multi-processing are key in the deployment of neuromorphic cores. 

\begin{table}[ht]
\centering
\caption{Individual parameters for our neurons, synapses, STDP learning rules, and simulation environment.}
\label{tab:params}
\begin{tabular}{cc}
\toprule
\textbf{Parameter}      & \textbf{Value} \\
\midrule
Paper to Paper $w$      & 100            \\
Paper to Topic $w$      & 1              \\
Val. to Topic $\tau_{-} $ & 30             \\
Topic Leak              & Off      \\
Paper Leak              & Off            \\
Val. / Test. Refractory & On             \\
Train/Topic Refractory  & Off           \\
Time Steps & 20 \\ 
\bottomrule
\end{tabular}
\end{table}
To evaluate the performance of Lava against NEST, we simulate both spiking graph implementations using the same parameters shown in Table \ref{tab:params}. All of paper and topic neurons have leak disabled. The initial paper-to-paper and train-to-topic weights are 100 and 1, respectively. The NEST implementation assumes $\tau_+ = \tau_- = 30$ so we configured our Lava implementation to have the same specification. The initial weights for the learnable synapses (test and validation to topic) vary from one implementation to another. In NEST, the initial weight is 0.0001 with floating-point integer precision. Since our implementation in Lava operates using fixed-point integer precision, we initialized the weights to 0 for our experiments to provide a fair comparison between the two simulators. 
Following the implementation proposed in \cite{cong2022semi}, the training and topic neurons are configured to have no refractory period while the testing and validation neurons have a refractory period greater than the number of simulation steps. By imposing a high refractory period on validation and testing neurons, we create temporal boundaries that limit their firing frequency where leakage does not affect the behaviour.
Meanwhile, the topic and training neurons operate without leakage or refractory periods to encourage continuous, reinforcing spiking activity within the learned synaptic connections. 


Similar to our method, the spiking graph in NEST infers the predicted topic as the topic neuron with the lowest reverse synaptic connection between the given paper and topics.
\renewcommand{\arraystretch}{1.25}

\begin{table}[h]
\centering
\caption{Comparing the validation classification accuracy and time per simulation step of NEST and Lava across varying time steps and delay periods.}
\label{tab:res_lava}
\begin{tabular}{c|cccc}
\toprule
\textit{\textbf{Simulator}}    & \multicolumn{2}{c}{\textbf{Lava}~\cite{lava}} & \multicolumn{2}{c}{\textbf{NEST}~\cite{Gewaltig:NEST}} \\ \hline
\textit{\textbf{Delay Period}} & 0        & 1               & 0               & 1               \\ \hline
$t=8$     & 55.57\%  & 28.57\% & 10.71\% & 70.71\% \\
$t=14$    & \textbf{63.57}\%  & 57.85\% & 12.14\% & \textbf{68.57}\% \\
$t=20$    & 36.43\%  & \textbf{59.28}\% & 12.14\% & \textbf{59.28}\% \\ \hline
$\Delta t(t_i)$ & 9.4ms  & 83.5ms & 1.4ms & 1.6ms \\ \bottomrule
\end{tabular}
\end{table}



Unless otherwise noted, all experiments are conducted with an Intel(R) Xeon(R) W-2295 CPU and 128GB of system memory. Table \ref{tab:res_lava} shows a quantitative comparison between Lava and NEST. The ideal configuration for our system, achieving 63.57\% accuracy on the validation set, used 14 simulation steps per paper with a delay of 0.
The original configuration from~\cite{cong2022semi} achieved 68.57\% accuracy with the same number of time steps and a delay period of 1. Reducing the delay period below 1 results in drastic performance reductions within the NEST implementation. Based on these results, our Lava implementation appears to exhibit greater resilience and robustness when delay values are modified compared to NEST.

NEST requires 1.6 milliseconds per simulation time step compared to Lava with 9.4 milliseconds. Despite NEST having a lower per-timestep latency, the entire graph has to be reinitialized for every paper, so that the network is reset to the initial inactive state. With this operation taking 9.73 seconds and being repeated for each of the 140 neurons in the validation set, it drastically increases the overall runtime.

In our Lava implementation, we reset the system by allowing it to discharge for 7 timesteps (i.e. the reset period) and re-initialize learnable weights to their initial values. Because of our custom neuron model, our network is only initialized once with a latency of 0.16 seconds and the reset process per paper takes 123.6 milliseconds.

It should also be noted that NEST has been heavily optimized for von-Neumann processors, whereas Lava is designed to be an abstract tool for prototyping biologically inspired systems before deployment to true neuromorphic hardware.

\begin{table}[]
\caption{The parameter search space for optimizing our fixed-point spiking graph neural network with Lava BO~\cite{10.1145/3589737.3605998}. The overall space contains 2450 parameters.}
    \label{tab:bo-search-space}
    \begin{tabular}{c|c|c}
    \hline
    \textit{\textbf{Parameter}} & \textit{\textbf{Options}}     & \textit{\textbf{Best}} \\ \hline
    Paper to Paper Weight              &  \{1, 50, 100, 300, 500, 700, 1000\}   & 300 \\
    Train to Topic Weight              &  \{1, 50, 100, 300, 500, 700, 1000\}   & 1   \\
    Val. to Topic $\tau_+$ \& $\tau_-$ &  \{20, 25, 30, 35, 40\}                & 30  \\
    Simulation Steps                   &  \{11, 12, ..., 20\}                   & 12  \\ \hline
    \end{tabular}
\end{table}
We use Bayesian optimization, implemented in Lava Bayesian Optimization (Lava BO)~\cite{10.1145/3589737.3605998}, to learn the optimal hyperparameter combination that maximizes classification accuracy with our spiking graph neural network. Our search space, as shown in Table~\ref{tab:bo-search-space}, consists 2450 unique parameter combinations across four individual parameters: paper to paper weight, topic to paper weight, $\tau_+$ and $\tau_-$, along with the number of simulation time steps. We provide Lava BO with 5 initial random points and allow it 10 iterations, for a total of 15 optimization iterations, to learn from this prior knowledge and intelligently select which individual parameter combinations could have provide the greatest increases in accuracy. This experiment concluded with a final parameter configuration providing performance increases of 3.58\% over the original NEST parameters for a total classification accuracy of 62.86\%. While this performance is not better than the optimized floating-point implementation, these results show that fixed precision spiking neural networks can be optimized with Bayesian optimization implemented in Lava BO~\cite{10.1145/3589737.3605998}.




\section{Conclusion}
In this work, we introduce a Loihi 2 compatible, fixed-precision spiking graph neural network for citation graph classification in Lava. Comparing against previous results from \cite{cong2022semi,cong2023}, our method achieves comparable accuracy while limiting inter-network dynamics to integer precision values. While the per-timestep simulation latency of our network is 5.31x slower compared with the highly optimized CPU implementation in Nest~\cite{Gewaltig:NEST}, our implementation provides native support for deployment to event-driven neuromorphic processors such as Loihi 2. Our spiking graph doesn't need to be fully reinitialized on a paper-by-paper basis, saving 9.73 seconds per paper evaluation. Moreover, we create a novel runtime process wrapper compatible with Lava Bayesian Optimization (Lava BO)~\cite{10.1145/3589737.3605998}. Using Lava BO, we were able to increase accuracy by 3.58\% over the original configuration from~\cite{cong2022semi} to a final value of 62.86\%.
While these results are not better than the optimized floating point implementation, they highlight the capability of fixed-precision, Loihi-compatible spiking neural networks to be optimized with Bayesian optimization.

In future works, we will deploy this architecture on physical Loihi 2 hardware.
Furthermore, we will use the improved runtime efficiency of our spiking spiking graph to improve hyperparameter optimization speed by running multiple instances of the spiking graph communicating with a single instance of Lava BO.
Lastly, we will experiment with different datasets and compare the latency, throughput, and power efficiency versus existing spiking neural network and graph convolution neural network implementations.
\begin{acks}
The work in this paper is supported by a gift from Intel Corporation.
\end{acks}

\bibliographystyle{ACM-Reference-Format}
\bibliography{sample-base}

\end{document}